\documentclass[
aps,
reprint,
prl,
superscriptaddress,
amsmath,
amssymb,
floatfix,
longbibliography,
noeprint
]{revtex4-2}

\usepackage{comment}
\usepackage{graphicx}
\usepackage[dvipsnames]{xcolor}
\usepackage[normalem]{ulem}
\usepackage{appendix}
\usepackage{silence}
\usepackage[T1]{fontenc}
\usepackage{currvita}

\definecolor{orange}{rgb}{1,0.5,0}
\definecolor{goodgreen}{rgb}{0.1,0.5,0}
\definecolor{goodred}{rgb}{0.7,0,0}
\usepackage{lineno}

\usepackage[colorlinks,urlcolor=goodgreen,citecolor=blue,linkcolor=goodred]{hyperref}

\usepackage{float}
\graphicspath{ {figures/} }

\usepackage{babel}

\begin{document}

\title{Perfect Spin-to-Charge Signal Conversion via Inverse Spin Hall Effect in a Superconductor}
\newcommand{\orcid}[1]{\href{https://orcid.org/#1}{\includegraphics[width=8pt]{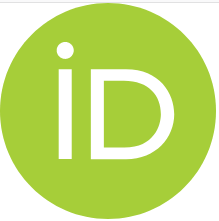}}}

\author{Rodrigo de las Heras\orcid{0009-0003-3640-1498}}
\email{rodrigo.delasheras@ehu.eus}
\affiliation{Centro de Física de Materiales (CFM-MPC) Centro Mixto CSIC-UPV/EHU, E-20018 Donostia-San Sebastián,  Spain}
\affiliation{Universidad del País Vasco (UPV/EHU)}

\author{Ilya V. Tokatly\orcid{0000-0001-6288-0689}}
\affiliation{Donostia International Physics Center (DIPC), 20018 Donostia-San
Sebastián, Spain}
\affiliation{Departamento de Polimeros y Materiales
Avanzados, Universidad del Pais Vasco UPV/EHU, 20018 Donostia-San Sebastian,
Spain}
\affiliation{IKERBASQUE, Basque Foundation for Science, 48009 Bilbao, Spain}

\author{F. Sebastian Bergeret\orcid{0000-0001-6007-4878}}
\affiliation{Centro de Física de Materiales (CFM-MPC) Centro Mixto CSIC-UPV/EHU, E-20018 Donostia-San Sebastián,  Spain}
\affiliation{Donostia International Physics Center (DIPC), 20018 Donostia-San Sebastián, Spain}

\begin{abstract}
Spin-to-charge conversion is central to spintronics, but conventional spin Hall mechanisms are intrinsically limited by the small spin Hall angle. Here we show that superconductivity can overcome this limitation and enable near-unity spin-to-charge signal conversion. We consider a nonlocal spin-valve geometry in which a spin accumulation is injected into a superconductor and converted into a transverse charge signal through the inverse spin Hall effect. In the superconducting state, the strongly suppressed quasiparticle conductance below the gap allows the resulting charge imbalance to generate a voltage comparable to the injected spin bias, despite a small spin Hall angle. We derive the full nonequilibrium response including finite spectral broadening and finite temperature, and propose temperature- and spin-bias-dependent measurements of the detector voltage to test our predictions.
\end{abstract}
\maketitle
The study of spin injection and accumulation in superconductors has a long history, encompassing a variety of conceptually distinct experiments \cite{PhysRevLett.26.192,PhysRevB.7.318,PhysRevB.66.140507,PhysRevLett.100.136601,yang2010extremely,PhysRevLett.109.207001,quay2013spin,PhysRevLett.112.036602,wakamura2015quasiparticle, Beckmann_2016}.   Spin injection into superconductors has emerged as a central topic in superconducting spintronics \cite{linder2015superconducting}, where dissipationless charge transport coexists with spin-dependent nonequilibrium phenomena.

In systems with strong spin–orbit coupling, spin injection can generate transverse charge currents via the inverse spin Hall effect (ISHE) \cite{PhysRevLett.85.393,omori2014inverse}, enabling electrical detection of spin transport. While the ISHE is well established in normal metals, its manifestation in superconductors is considerably more subtle. Superconductivity profoundly modifies the quasiparticle spectrum, and hence, the relaxation characteristic length, leading to qualitatively distinct nonequilibrium responses compared to the normal state.

Previous theoretical studies of spin injection in superconductors \cite{aronov1976spin,PhysRevB.67.052409,PhysRevB.70.212508,RevModPhys.90.041001}, and consequently of the ISHE in superconductors \cite{Takahashi_2012,PhysRevB.95.054509}, have generally neglected processes occurring within a distance of order $\xi$ from the spin injector, where $\xi$ is the superconducting coherence length. They have also typically assumed that the injector, usually a normal metal or a ferromagnet, does not perturb the superconducting spectrum.
While such approximations can be appropriate for charge injection \cite{PhysRevLett.28.1363,PhysRevLett.28.1366,PhysRevB.81.184524}, they need not hold for spin injection. As recently demonstrated in Ref.~\cite{tokatly2025spingalvanicresponsenonequilibriumspin}, neglecting the local modification of superconductivity can miss essential contributions to the nonequilibrium response and lead to an incomplete interpretation of spin-injection experiments. How such contributions affect the inverse spin Hall  voltage measured
in a nonlocal geometry remains an open question. In particular, it is not
clear whether superconductivity suppresses or enhances the detected voltage
relative to the normal state, nor which experimental signatures would
unambiguously reveal these effects.

In this Letter, we address these questions. We develop a theory of
 ISHE in superconductors.  We find that,
contrary to the naive expectation of a signal suppressed by the opening of
the gap, superconductivity can strongly enhance the detected voltage.
 Remarkably, in an ideal BCS superconductor, as $T \to 0$, the spin-to-charge signal
conversion efficiency approaches unity, enabling near-perfect conversion of
spin accumulation into a charge voltage. Finite spectral broadening due to inelastic processes, and
temperature reduce this efficiency. Nevertheless, over a wide range of temperatures and spin biases, the detected voltage remains orders of
magnitude larger than in the normal state.

\begin{figure}[tb]
    \centering
    \includegraphics[width=\linewidth]{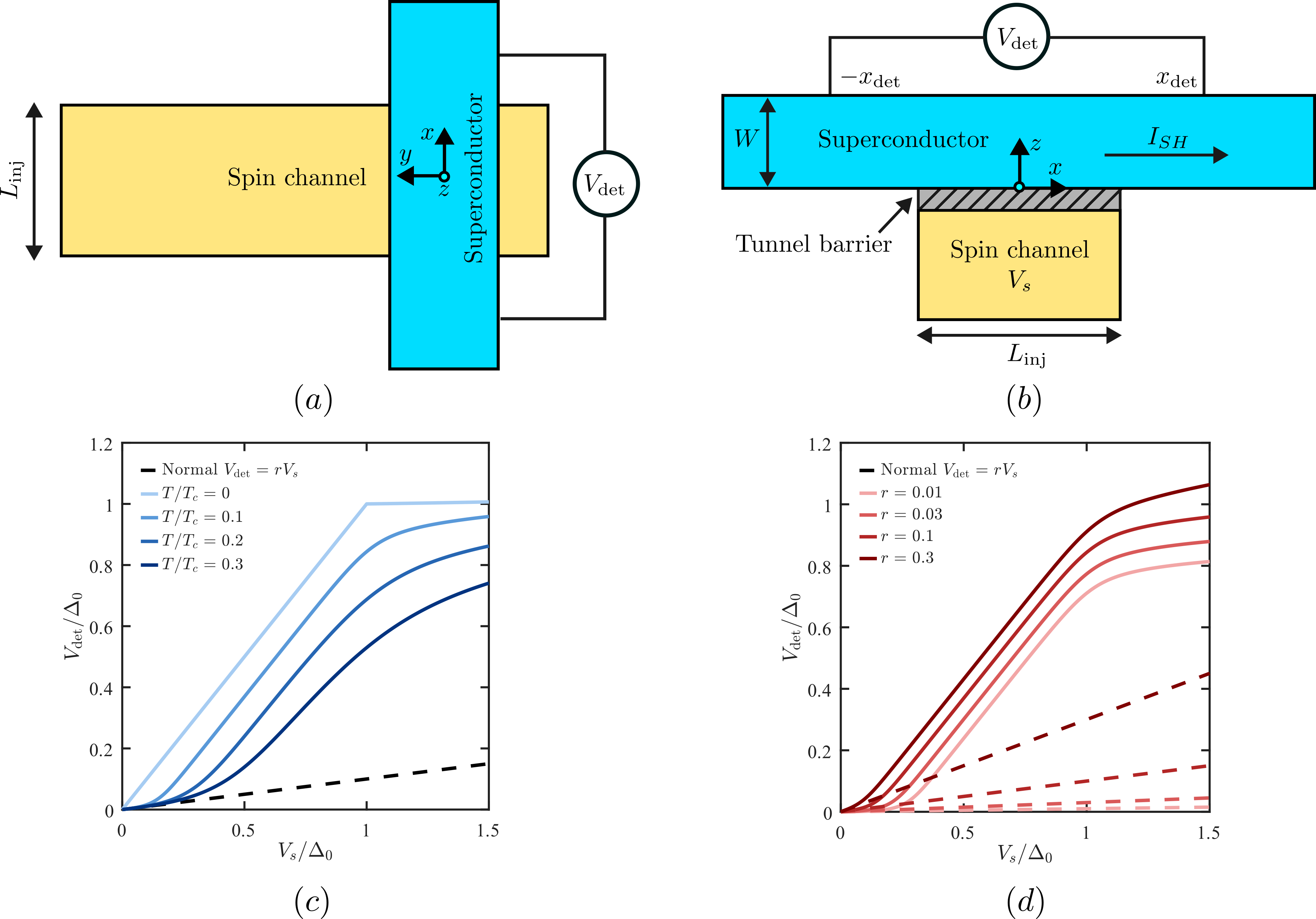}
    \caption{(a)--(b) Schematic illustrations of the setup considered in this work. A two-dimensional superconducting wire of width $W$ is coupled through a tunnel barrier to a normal-metal spin channel of length $L_{\mathrm{inj}}$ ($x\in[-\frac{L_{\mathrm{inj}}}{2},\frac{L_{\mathrm{inj}}}{2}]$) at spin bias $V_s$. A normal-metal voltage probe is positioned at a distance $x_{\mathrm{det}}$ from the center of the spin channel. (c)--(d) Detector voltage $V_\mathrm{det}$ as a function of the spin bias for a perfect BCS superconductor. The superconducting gap is taken to follow the temperature dependence $\Delta=\Delta_0\tanh\left(1.74\sqrt{T_c/T-1}\right)$. (c) For $r=0.1$ and several values of $T$. (d) For $T/T_c=0.1$ and several values of $r$. The dashed lines show the normal state behavior $V_{\rm det}=r V_s$}
    \label{fig:Setup}
\end{figure}

Specifically, we consider the typical nonlocal spin-valve geometry, sketched  in Fig.~\ref{fig:Setup}\textcolor{goodred}{(a)--(b)}. The normal-metal channel supports a nonequilibrium spin accumulation characterized by a spin bias $V_s$, defined as the difference between electrochemical potentials of spin species \cite{PhysRevB.67.085319,RevModPhys.90.041001}. Along the interface of length $L_{\mathrm{inj}}$, part of this spin accumulation is injected into the superconductor, where spin-to-charge conversion mediated by the spin Hall angle $\theta$ generates a transverse charge response. This response is detected as a voltage $V_{\mathrm{det}}$ using a normal-metal probe.

We first focus on the well known normal-state result. For the geometry of Fig.~\ref{fig:Setup}\textcolor{goodred}{(a)--(b)} the spin accumulation $S$ injected into the superconductor satisfies the spin-diffusion equation and. If $L_{\mathrm{inj}}$  is much larger than the spin relaxation length $\kappa_s^{-1}$,  deep within the contact region, the spin injected  varies across the thickness $z\in[0,W]$ as
\begin{equation}\label{eq:spin_normal}
    S(z)=N_0V_s \frac{R_s}{R_b}\frac{\cosh(\kappa_s(z-W))}{\sinh(\kappa_sW)}.
\end{equation}
where  $R_b$ is the normal-state interface resistance, and
$R_s=1/(\kappa_s\sigma_n A)$ is the spin resistance, with $\sigma_n$ the normal-state Drude conductivity,  $A$ the contact area, and $N_0$ the density of states at the  Fermi level.
The spin accumulation gradient drives a spin current, which is converted by the inverse spin Hall effect into a transverse charge current (the spin Hall current). Under open-circuit conditions, a transverse voltage $V_{\mathrm{det}}$ is generated, given by:
\begin{equation}\label{eq:Vdet_normal}
    V_{\mathrm{det}} = \theta\frac{L_{\mathrm{inj}}}{W}\frac{S(0)-S(W)}{N_0}.
\end{equation}
Combining Eqs.~\eqref{eq:spin_normal} and \eqref{eq:Vdet_normal}, one obtains:
\begin{equation}\label{eq:VrV}
    V_{\mathrm{det}}=rV_s
\end{equation}
with the normal-state spin-to-charge-signal conversion efficiency:
\begin{equation}
    r=\theta \frac{L_{\mathrm{inj}}}{W}\frac{R_s}{R_b}\tanh\left(\frac{\kappa_s W}{2}\right).
\end{equation}
Thus, in the normal state, $r$ is limited by the small spin Hall angle $\theta$, which is typically only a few percent.

The superconducting state is qualitatively different. The injected spin accumulation is converted into a quasiparticle charge imbalance $Q^*$, which sets the electrochemical potential of the quasiparticles relative to that of the condensate. The measured voltage, $V_{\rm det}$, Fig.~\ref{fig:Setup}\textcolor{goodred}{(a)--(b)}, is determined  
by the  difference of the charge imbalance across the detector $\delta Q^*$ via  \cite{PhysRevLett.28.1366,RevModPhys.90.041001}
\begin{equation}\label{eq:chargedef}
    \delta Q^*=\int_0^{\infty} N_\omega f_T(V_{\mathrm{det}})d\omega
\end{equation}
where
\begin{equation}
N_\omega=\operatorname{Re}\left(\frac{\omega}{\sqrt{\omega^2-\Delta^2}}\right)
\end{equation}
is the normalized BCS density of states, 
\begin{equation}
    f_T(V)=n_{FD}(\omega-V)-n_{FD}(\omega+V)\; , 
\end{equation}
and $n_{FD}$ is the Fermi-Dirac distribution.

The charge-imbalance drop generated by the spin bias $V_s$ at the contact with the spin channel, is given by (see Appendix \ref{B3} for details):
\begin{equation}
    \delta Q^*=r\int_\Delta^{\infty} N_\omega f_T(V_s)d\omega.
\end{equation}
Combining this expression with Eq.~\eqref{eq:chargedef} yields an integral equation for the detector voltage:
\begin{equation}\label{eq:integral}
    \int_\Delta^{\infty} N_\omega f_T(V_{\mathrm{det}})d\omega=r\int_\Delta^{\infty} N_\omega f_T(V_s)d\omega.
\end{equation}

In  the high-temperature limit $T\gg V_s$, this equation leads to the normal state result, Eq.~(\ref{eq:spin_normal}), see  
Appendix~\ref{ap:A1}. Whereas in the limit $T\to0$, gives (see Appendix~\ref{ap:A3} for details):
\begin{equation}\label{eq:T0limit}
\lim_{T\to0}V_{\mathrm{det}}=\begin{cases}
 V_s  & \text{if }  V_s <\Delta_0 \\
 \sqrt{\Delta_0^2+r^2(V_s^2-\Delta_0^2)} & \text{if }  V_s>\Delta_0 
\end{cases}.
\end{equation}
Thus, at a zero temperature, for spin biases below the superconducting gap, the spin-to-charge signal conversion efficiency is independent of the spin Hall angle $\theta$ and 
reaches unity. 

At finite temperature, the conversion efficiency falls below this ideal
limit, but over a wide range of temperatures it nevertheless remains well
above its normal-state value.
Figures~\ref{fig:Setup}\textcolor{goodred}{(c)--(d)} show the  $V_{\rm det}(V_s)$ dependence by numerically solving Eq.~\eqref{eq:integral}. At finite temperatures, three regimes emerge: (i) an initial linear regime with slope $r$, as in the normal state; (ii) an intermediate regime in which $V_{\mathrm{det}}$ approaches $V_s$, corresponding to nearly unit conversion efficiency; and (iii) a high-bias regime in which the response again becomes linear with slope $r$. The crossover into the enhanced regime occurs at a spin bias of order $V_s\sim T|\ln r|$.
 
Equation~\eqref{eq:integral} is valid for an ideal BCS superconductor. In real superconductors though, inelastic scattering processes are unavoidable, even at $T\rightarrow
0$ \cite{PhysRevLett.105.026803,PhysRevB.102.134506}. Such inelastic processes, described usually by the Dynes parameter $\eta$, leads to a finite density of states at zero energy of the order of $\eta$. 
In the remainder of this Letter, we generalize  Eq.~(\ref{eq:integral}) for finite $\eta$ and show that, even though $V_{\rm det}$ is reduced by inelastic processes,  the signal can remain orders of magnitude larger than in the normal state.

Our starting point is the Usadel equation, generalized to include spin Hall effect
\cite{PhysRevB.104.064515,PhysRevB.94.180502,PhysRevB.98.144515}:
\begin{equation}\label{eq:Usadel}
   \partial_k\mathcal{J}_k+[-i\omega\tau_3+\Delta e^{i\tau_3\varphi}\tau_1,g]+\frac{1}{8\tau_{so}}[\sigma_ig\sigma_i,g]=\mathcal{T}
\end{equation}
here $g$ denotes the quasiclassical Green's function matrix (GF) in the combined Keldysh-Nambu-spin space, which satisfies the normalization condition $g^2=1$ \cite{RevModPhys.90.041001}. The symbols $\sigma_i$ and $\tau_i$ denote Pauli matrices acting in spin and Nambu space, respectively. Furthermore, $\varphi$ is the superconducting phase and $\tau_{so}$ is the spin-relaxation time, which we assume originates from spin-orbit interaction. The matrix current $\mathcal{J}_k$ and torque $\mathcal{T}$ are given by:  
\begin{equation}
    \mathcal{J}_k=-D\left(g\partial_kg-\epsilon_{klm}\frac{\theta}{4}\{ \partial_lg,\sigma_m+g\sigma_mg\}\right)
\end{equation}
\begin{equation}
    \mathcal{T}=\frac{D}{4}\epsilon_{ijk}\theta[\sigma_k,g \partial_ig \partial_jg]
\end{equation}
where $D$ is the normal state diffusion constant, $\epsilon_{ijk}$ the Levi-Civita tensor and $\theta$ the spin-hall angle.

To describe spin injection from a normal metal electrode coupled to the superconductor through a tunnel barrier, we employ the generalized Kupriyanov-Lukichev boundary conditions \cite{kuprianov1988influence}:
\begin{equation}\label{eq:KLbcond}
    n_k\mathcal{J}_k=\frac{D}{2\sigma_nR_bA}[g,g_{\mathrm{inj}}]
\end{equation}
where $n_k$ is a unit vector normal to the interface. The GF at the injector, $g_{\mathrm{inj}}$, has the retarded, advanced, and Keldysh components $g_{\mathrm{inj}}^R=-g^A_{\mathrm{inj}}=\tau_3$ and $g_{\mathrm{inj}}^K=2\tau_3\tanh\left(\frac{\omega+\sigma_y V_s}{2T}\right)$, respectively.

Because the Usadel equation is highly nonlinear, it generally must be solved numerically alongside the self-consistency equation for $\Delta$. However, expanding perturbatively in the inverse interface resistance, $R_b^{-1}$, allows for analytical solutions in the tunnel limit. This limit also offers the clearest interpretation of experiments, because highly transparent interfaces induce pronounced proximity effects that suppress the gap and obscure intrinsic physics of the superconductor.

{\it Spin injection.}  For the spin injection process, we set $\theta=0$. In this way, the spin accumulation is determined to zeroth order in $\theta$, while the charge response generated by this spin accumulation will appear at first order in $\theta$. Our treatment therefore corresponds to a perturbative expansion in the spin Hall angle.

Under these assumptions, we show in Appendix \ref{B1} that the spatial profile of the injected spin accumulation within the contact region ($\vert{}x\vert{} \ll L_{\mathrm{inj}}/2$) is \cite{RevModPhys.90.041001}:
\begin{equation}\label{eq:spin_superconductor}
    S(z)=\frac{N_0}{8}\int_0^\infty\operatorname{Tr}\left(\sigma_y\tau_3g^K\right)d\omega=\int_0^{\infty} S_{\omega}(z)d\omega
\end{equation}
where the spectral spin accumulation per quasiparticle energy $\omega$ reads:
\begin{widetext}
\begin{equation}\label{eq:S_omega}
    S_\omega(z)=N_0\frac{R_s}{2R_b}\left(1+\frac{\Delta^2+\omega^2+\eta^2}{|\Delta^2-(\omega+i\eta)^2|}\right)\frac{f_T(V_s)}{\sqrt{1+\frac{\kappa_\omega^2}{\kappa_s^2}}}\frac{\cosh\left(\sqrt{\kappa_s^2+\kappa_\omega^2}(z-W)\right)}{\sinh\left(\sqrt{\kappa_s^2+\kappa_\omega^2}W\right)}
\end{equation}
\end{widetext}
and $\kappa_\omega =\sqrt{\frac{2}{D}\operatorname{Re}\left(\sqrt{\Delta^2-(\omega+i\eta)^2}\right)}$
defines the energy-dependent decay length into the superconductor.

Remarkably, $S_\omega(z)$ remains finite even for subgap energies ($\vert{}\omega\vert{}<\Delta$), a behavior discussed in detail in Ref.~\cite{tokatly2025spingalvanicresponsenonequilibriumspin}.
For $\Delta,\kappa_\omega\to0$ in Eqs.~\eqref{eq:spin_superconductor} and \eqref{eq:S_omega} straightforwardly recovers the normal-state result in Eq.~\eqref{eq:spin_normal}. 

{\it Spin to charge conversion.}
Via the inverse spin Hall effect, the gradient of this spin accumulation gives rise to a transverse charge current. Under open-circuit conditions, this current is screened by two distinct contributions: a Cooper-pair supercurrent and a quasiparticle current that induces a charge imbalance $Q^*$. In Appendix \ref{B2}, we show that the total charge-imbalance drop across the detector, $\delta Q^*$, generated by the spin accumulation in Eq.~\eqref{eq:S_omega} is given by:
\begin{equation}\label{eq:deltaQ}
    \delta Q^*=2\theta\int_0^\infty  \frac{q_\omega (S_\omega(0)-S_\omega(W))}{N_0\kappa_\omega W}\sinh\left(\frac{\kappa_\omega L_{\mathrm{inj}}}{2}\right)e^{-\kappa_\omega x}d\omega
\end{equation}
where $q_\omega=\operatorname{Re}\left(\frac{\sqrt{(\omega+i\eta)^2-\Delta^2}}{\omega}\right)$ is the effective quasiparticle charge \cite{tinkham2004introduction}. Similarly to the normal state, charge imbalance remains proportional to the spin-accumulation drop across the two interfaces, now weighted by the quasiparticle charge $q_\omega$. Setting $\Delta,\kappa_\omega\rightarrow0$ in Eq.~\eqref{eq:deltaQ} yields $\delta Q^*=V_{\mathrm{det}}=rV_s$, recovering the normal state result from Eq.~\eqref{eq:VrV}.

Substituting the result from Eq.~\eqref{eq:deltaQ} in Eq.~\eqref{eq:chargedef} yields:
\begin{widetext}
\begin{equation}\label{eq:integraldynes}
    \int_0^{\infty} N_\omega f_T(V_{\mathrm{det}})d\omega=    2\theta\int_0^\infty  \frac{q_\omega (S_\omega(0)-S_\omega(W))}{N_0\kappa_\omega W}\sinh\left(\frac{\kappa_\omega L_{\mathrm{inj}}}{2}\right)e^{-\kappa_\omega x_{\mathrm{det}}}d\omega.
\end{equation}
\end{widetext}
Here, the density of states includes Dynes broadening and is given by
\begin{equation}
    N_\omega=\operatorname{Re}\left(\frac{\omega+i\eta}{\sqrt{(\omega+i\eta)^2-\Delta^2}}\right).
\end{equation}
In the limit $\eta\to0$, Eq. (\ref{eq:integraldynes})  reduces to Eq.~\eqref{eq:integral}, recovering the ideal-BCS result.

\begin{figure}[tb]
    \centering
    \includegraphics[width=\linewidth]{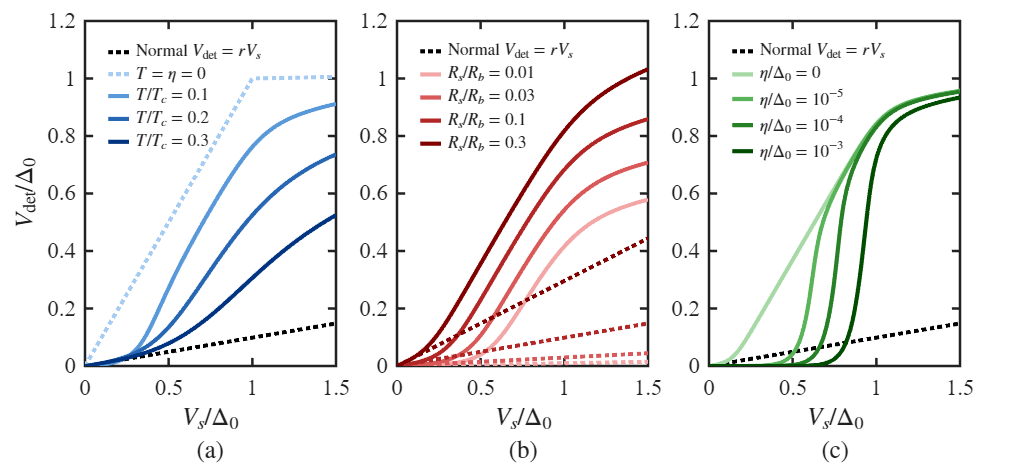}
    \caption{Detector voltage $V_{\mathrm{det}}$ as a function of the spin bias $V_s$ for $L_{\mathrm{inj}}/\xi_0=50$, $W/\xi_0=5$, $\kappa_s\xi_0=1$, $x_{\mathrm{det}}/\xi_0=100$ and $\theta=0.1$. Here $\xi_0=\sqrt{D/\Delta_0}$ is the zero temperature superconducting coherence length  The superconducting gap is taken to follow the temperature dependence $\Delta=\Delta_0\tanh\left(1.74\sqrt{T_c/T-1}\right)$. (a) For $\eta/\Delta_0=10^{-4}$ and $R_s/R_b=0.1$ for several values of $T$. (b) For $T/T_c=0.2$ and $\eta/\Delta_0=10^{-4}$ for several values of $R_s/R_b$. (c) For $T/T_c=0.1$ and $R_s/R_b=0.1$ for several values of $\eta$. A strong nonlinear enhancement of the detector signal develops at low temperatures when the injected spin accumulation approaches the superconducting gap $V_s \sim \Delta_0$. The dashed lines show the normal state behavior $V_{\rm det}=r V_s$}
    \label{fig:HugeSignal}
\end{figure}

In figure \ref{fig:HugeSignal} we show the $V_{\rm det}(V_s)$ dependence by numerically solving Eq.~\eqref{eq:integraldynes}. 
Interestingly, for a large range of subgap spin voltages, $V_{\rm det}$ is larger than the voltage detected in the normal state (dashed lines), in agreement with the analytical results discussed in Fig.~\ref{fig:Setup}. A  particular pronounced nonlinear enhancement of $V_{\rm det}$ emerges at low temperatures as $V_s$ approaches the superconducting gap.  From panels (b) and (c) from Fig.~\ref{fig:HugeSignal} it is evident that both increasing temperature $T$ or the broadening $\eta$ affect very substantially to the amount of signal. These results suggest two complementary experimental tests. First, at fixed
temperature well below $T_c$, one may increase the spin bias $V_s$. While in
the normal state $V_{\rm det}$ grows linearly with $V_s$, in the
superconducting state it grows strongly nonlinearly: beyond a threshold spin
bias, $V_{\rm det}$ exceeds its normal-state value, and the enhancement grows
rapidly, reaching in certain cases one to two orders of magnitude as $V_s$ approaches
$\Delta_0$. Second, at fixed and sufficiently large spin bias, one may measure
$V_{\rm det}$ while cooling the sample from the normal into the
superconducting state. $V_{\rm det}$ should then rise well above its
normal-state value,  before
decreasing at the lowest temperatures.

{\it Closed loop geometry.} For a wire closed into a ring, i.e., periodic along the $x$ direction (the same geometry as in Ref. \cite{omori2014inverse}), the total circulating spin-Hall current is (see Appendix \ref{B1}):
\begin{equation}\label{eq:spinhallcurrent}
I_{SH}
=
-\theta
\frac{\sigma_n}{e}\frac{L_{\mathrm{inj}}}{L}\int_0^\infty\frac{S_\omega(0)-S_\omega(W)}{N_0}d\omega
\end{equation}
where $e$ the electron charge and $L$ the length of the loop. As discussed above, this current is generated by the transverse gradient of the spin accumulation across the wire width $W$.

For a subgap spin bias, $V_s<\Delta$, the dominant contribution to Eq.~\eqref{eq:spinhallcurrent} comes from subgap energies. Consequently, within one coherence length it will be converted into supercurrent. This supercurrent could be detected through its magnetic flux, for example using a SQUID. For $V_s>\Delta$, in contrast, quasiparticle contributions become important and generate a charge imbalance $Q^*$. At distances shorter than the charge-imbalance relaxation length $\lambda_{Q^*}$, this nonequilibrium current can therefore be detected with a normal-metal voltage probe. At distances larger than $\lambda_{Q^*}$, the quasiparticle contribution relaxes and the current is progressively transferred to the condensate, recovering a supercurrent.

{\it Conclusions.}
In conclusion, we have shown that superconductivity can dramatically enhance the conversion of a spin bias into a charge signal through the inverse spin Hall effect. In contrast to the normal state, where the conversion efficiency is directly limited by the spin Hall angle, the superconducting state allows to produce a voltage of the same order as the  spin bias. In the ideal BCS limit, this leads to $V_{\mathrm{det}}\approx V_s$ at low temperature for subgap injection, corresponding to unit spin-to-charge signal conversion efficiency even for a small spin Hall angle.

In a closed superconducting geometry, the same spin-Hall response generates a circulating current. For subgap injection this current is predominantly converted into a supercurrent, providing a magnetic-flux signature accessible to SQUID measurements. Our results thus identify superconducting spin Hall devices as a route toward efficient spin-to-charge signal conversion.

{\it Acknowledgments.}
R.~H. and F.~S.~B. thank financial support from the the Spanish MCIN/AEI/10.13039/501100011033 
through the grants PID2023-148225NB-C31, and TED2021-130292B-C41. F. S. B. also thanks financial support from the
European Union’s Horizon Europe research and innovation program under grant agreement No. 101130224 (JOSEPHINE). I.~V.~T. acknowledges support from the
Spanish MCIN/AEI/10.13039/501100011033 through
the Project No. PID2023-148225NB-C32, and the
Basque Government (Grant No. IT1453-22). 

\bibliography{biblio} 

\onecolumngrid
\appendix
\section{Analysis of the spin-Hall signal at low temperatures}\label{ap:lowTsignal}
For a perfect BCS superconductor, the detector voltage $V_\mathrm{det}$ is determined by Eq.~\eqref{eq:integral}. We begin by evaluating the integral
\begin{equation}\label{eq:A1}
    I(V)=\int_\Delta^\infty N_\omega f_T(V) d\omega.
\end{equation}
\subsection{High temperature limit}\label{ap:A1}
In this limit, the distribution function can be expanded in powers of $V/T$. Using
\begin{equation}
    f_T(V)=n_{FD}(\omega-V)-n_{FD}(\omega+V)\simeq-2V\frac{\partial n_{FD}(\omega)}{\partial\omega}+\mathcal{O}\left(\frac{V^2}{T^2}\right)
\end{equation}
the integral becomes:
\begin{equation}
    I(V)\simeq VY(T)
\end{equation}
where $Y(T)$ is the Yoshida function:
\begin{equation}
    Y(T)=2\int_\Delta^\infty \frac{\omega}{\sqrt{\omega^2-\Delta^2}}\left(-\frac{\partial n_{FD}(\omega)}{\partial\omega}\right)d\omega.
\end{equation}
Substitution into Eq.~\eqref{eq:integral} then gives, to leading order in $V/T$,
\begin{equation}
    V_{\mathrm{det}}=rV_s.
\end{equation}

\subsection{Low temperature limit}\label{ap:A2}
In this limit, it is convenient to rewrite the distribution function as
\begin{equation}
    f_T(V)=\frac{\sinh\frac{V}{T}}{\cosh\frac{\omega}{T} + \cosh\frac{V}{T}}=\frac{2e^{-\omega/T}\sinh\frac{V}{T}}{1+2e^{-\omega/T}\cosh\frac{V}{T}+e^{-2\omega/T}} \simeq 2e^{-\omega/T}\sinh\frac{V}{T}\left(1+O\left(e^{-\frac{\omega-V}{T}}\right)\right),
\end{equation}
where the last step is valid if $e^{-\frac{\Delta-V}{T}}\ll1$, because the lower integration limit is $\omega=\Delta$. Substituting this expansion, we obtain
\begin{equation}
    {I}(V)= 2\Delta\sinh\frac{V}{T}K_1\left(\frac{\Delta}{T}\right)
\end{equation}
where $K_1(x)$ is the modified Bessel function of second kind. Using this result, Eq.~\eqref{eq:integral} simplifies as
\begin{equation} \label{V_d-equation-1}
    \sinh\frac{V_\mathrm{det}}{T}=r\sinh\frac{V_s}{T}.
\end{equation}
The solution to this equation reads:
\begin{equation}\label{eq:A9}
    V_{\mathrm{det}} = T\ln\left(r\sinh\frac{V_s}{T} + \sqrt{1+r^2\,\sinh^2\frac{V_s}{T}}\, \right).
\end{equation}
It is clear that the parameter which controls the function from Eq.~\eqref{eq:A9} is $r\sinh(V_s/T)$. The limiting forms of the solution are:
\begin{equation}
    V_{\mathrm{det}}\simeq\begin{cases}
 rV_s  & \text{if } r\sinh\frac{V_s}{T}\ll1 \\
 V_s+T\ln r & \text{if } r\sinh\frac{V_s}{T}\gg1 
\end{cases}.
\end{equation}
These two limits describe a crossover from the regime
$V_{\mathrm{det}}\approx rV_s$, which has a normal-state-like behavior, to the regime
$V_{\mathrm{det}}\approx V_s$, where the detector voltage closely follows the spin bias. The crossover between these two regimes occurs when
$V_s\sim T|\ln r|$.

\subsection{Zero temperature limit}\label{ap:A3}
For $V_s<\Delta_0$, taking the limit $T\to0$ in Eq.~\eqref{eq:A9} yields
\begin{equation}
    V_{\rm det}=V_s\quad \mathrm{if} \quad V_s<\Delta_0.
\end{equation}
For $V_s>\Delta_0$, the integral appearing in Eq.~\eqref{eq:A1} can be evaluated exactly. In the zero-temperature limit, $f_T(V)$ becomes a step function, giving
\begin{equation}
    I(V)=\lim_{T\to0}\int_\Delta^\infty N_\omega f_T(V) d\omega=\int_{\Delta_0}^V N_\omega d\omega=\sqrt{V^2-\Delta_0^2}.
\end{equation}
Using this result we obtain
\begin{equation}
    V_{\rm det}=\sqrt{\Delta_0^2+r^2(V_s^2-\Delta_0^2)}\quad \mathrm{if} \quad V_s>\Delta_0.
\end{equation}
Combining the two regimes, the zero-temperature detector voltage is therefore
\begin{equation}
    \lim_{T\to0}V_{\mathrm{det}}=\begin{cases}
 V_s  & \text{if }  V_s\leq\Delta_0 \\
 \sqrt{\Delta_0^2+r^2(V_s^2-\Delta_0^2)} & \text{if }  V_s>\Delta_0 
\end{cases}.
\end{equation}
\section{Derivation of the Inverse Spin Hall effect in a Superconductor}\label{ap:derivation}
The retarded/advanced and Keldysh components of Eq.~\eqref{eq:Usadel} are given by:
\begin{equation}
    -D\partial_k(g^{R/A}\partial_kg^{R/A})+[-i(\omega\pm i\eta)\tau_3+\Delta e^{i\tau_3\varphi}\tau_1,g^{R/A}]+\frac{1}{8\tau_s}[\sigma_ig^{R/A}\sigma_i,g^{R/A}]=0
\end{equation}
\begin{equation}\label{eq:Usadel_Keldysh}
    -D\partial_k(g^R\partial_kg^K+g^K\partial_kg^A)+[-i\omega\tau_3+\Delta e^{i\tau_3 \varphi}\tau_1,g^K]+\frac{1}{8\tau_s}[\sigma_i,g^R\sigma_i g^K+g^K \sigma_i g^A]=0.
\end{equation}
To first order in the spin Hall angle $\theta$, the boundary conditions in Eq.~\eqref{eq:KLbcond} simplify to:
\begin{align}
    &-\left.g^{R/A}\partial_zg^{R/A}\right|_{z=0}=\pm\frac{1}{2\sigma_nR_bA}\Theta\left(\frac{L_{\mathrm{inj}}}{2}-|x|\right)[g^{R/A},\tau_3]
\\
    &-\left.g^{R/A}\partial_zg^{R/A}\right|_{z=W}=0
\end{align}
\begin{align}\label{eq:KLbcond_Keldysh1}
    &-\left.(g^R\partial_z g^K+g^K\partial_z g^A-\theta\sigma_y\partial_xg^K)\right|_{z=0}=\frac{1}{2\sigma_nR_bA}\Theta\left(\frac{L_{\mathrm{inj}}}{2}-|x|\right)\left[2\left(f_L(V_s)+f_T(V_s)\sigma_y\right)(g^R\tau_3-\tau_3g^A)-\{g^K,\tau_3\}\right]
\\ \label{eq:KLbcond_Keldysh2}
    &-\left.(g^R\partial_z g^K+g^K\partial_z g^A-\theta\sigma_y\partial_xg^K)\right|_{z=W}=0
\end{align}
where $\Theta(x)$ denotes the Heaviside step function and:
\begin{equation}
    f_L(V)=1-n_{FD}(\omega+V)-n_{FD}(\omega-V).
\end{equation}
Note that to linear order in $\theta$, the retarded and advanced Green's functions remain spin independent and thus contain only singlet components.

We next perform a perturbative expansion in the inverse interface resistance $R_b^{-1}$ (it will be shown later that the expansion is in reality controlled by the dimensionless parameter $R_s/R_b$, where $R_s$ is the spin resistance of the interface):
\begin{equation}\label{eq:gRA_expansion}
    g^{R/A}=g_0^{R/A}+g_{1s}^{R/A}+\mathcal O(R_b^{-2})
\end{equation}
\begin{equation}\label{eq:gK_expansion}
    g^{K}=g_0^K+(g_{1s}^R-g_{1s}^A)f_L(0)+ g_{1s}^K+\sigma_yg_{1t}^K+\mathcal O(R_b^{-2})
\end{equation}
where the numerical subscripts denote the order in $R_b^{-1}$, while $s$ and $t$ label singlet and triplet components, respectively. The zeroth-order Green's functions correspond to an equilibrium bulk BCS superconductor:
\begin{equation}
    g_0^{R/A}= \frac{-i(\omega\pm i\eta)\tau_3+\Delta\tau_1}{\sqrt{\Delta^2-(\omega\pm i\eta)^2}}
\end{equation}
\begin{equation}
    g_0^K=(g_0^R-g_0^A)f_L(0).
\end{equation}

\subsection{2D Spin injection in a superconductor}\label{B1}
Substituting Eqs.~\eqref{eq:gRA_expansion} and \eqref{eq:gK_expansion} into Eq.~\eqref{eq:Usadel_Keldysh}, retaining terms to first order in $R_b^{-1}$, and taking the trace over spin space with $\sigma_y$ yields the equation for the triplet component $g_{1t}^K$:
\begin{equation}
    \nabla^2g_{1t}^{K}-(\kappa_s^2+\kappa_\omega^2)g_{1t}^{K}=0
\end{equation}
where 
\begin{equation}
    \kappa_s = \sqrt{\frac{1}{\tau_sD}}, \qquad \kappa_\omega = \sqrt{\frac{2}{D}\operatorname{Re}\left(\sqrt{\Delta^2-(\omega+i\eta)^2}\right)}.
\end{equation}
The corresponding boundary conditions follow from Eqs.~\eqref{eq:KLbcond_Keldysh1} and \eqref{eq:KLbcond_Keldysh2}:
\begin{equation}
    \left.\partial_z g_{1t}^K\right|_{z=0}=-\frac{f_T(V_s)}{\sigma_nR_bA}\Theta\left(\frac{L_{\mathrm{inj}}}{2}-|x|\right)(\tau_3-g_0^R\tau_3g_0^A)
\end{equation}
\begin{equation}
    \left.\partial_z g_{1t}^K\right|_{z=W}=0.
\end{equation}
Solving via separation of variables yields:
\begin{align}\label{eq:B16}
    g_{1t}^K&=(\tau_3-g_0^R\tau_3g_0^A)\frac{f_T(V_s)}{\sigma_nR_bA}\int_{-\infty}^{\infty} \frac{\Theta_q}{\sqrt{\kappa_s^2+\kappa_\omega^2+q^2}}\frac{\cosh\left(\sqrt{\kappa_s^2+\kappa_\omega^2+q^2}(z-W)\right)}{\sinh\left(\sqrt{\kappa_s^2+\kappa_\omega^2+q^2}W\right)}\frac{e^{iqx}}{2\pi}dq\notag\\
    &\approx (\tau_3-g_0^R\tau_3g_0^A)\frac{f_T(V_s)}{\sqrt{\kappa_s^2+\kappa_\omega^2}\sigma_nR_bA}\frac{\cosh\left(\sqrt{\kappa_s^2+\kappa_\omega^2}(z-W)\right)}{\sinh\left(\sqrt{\kappa_s^2+\kappa_\omega^2}W\right)}\Theta\left(\frac{L_{\mathrm{inj}}}{2}-|x|\right)
\end{align}
where $\Theta_q=\frac{2}{q}\sin(\frac{qL_{\mathrm{inj}}}{2})$. The approximation on the second line holds in the limit $L_{\mathrm{inj}}\gg \mathrm{max}\left(\frac{1}{\sqrt{\kappa_s^2+\kappa_\omega^2}},W\right)$ and sufficiently far from the injector edges, i.e., $\left||x|-\frac{L_{\mathrm{inj}}}{2}\right|\gg\frac{1}{\sqrt{\kappa_s^2+\kappa_\omega^2}}$. These conditions are satisfied in the experimentally relevant regime where spin relaxation dominates.

The resulting spin accumulation is given by:
\begin{equation}
    S =\frac{N_0}{8}\int_0^\infty\operatorname{Tr}\left(\sigma_y\tau_3g^K\right)d\omega= \frac{N_0}{8}\int_{0}^\infty \operatorname{Tr}(\tau_3 g_{1t}^K)d\omega=\Theta\left(\frac{L_{\mathrm{inj}}}{2}-|x|\right)\int_0^\infty S_\omega(z)d\omega
\end{equation}
where:
\begin{equation}
    S_\omega(z)=N_0\frac{R_s}{2R_b}\left(1+\frac{\Delta^2+\omega^2+\eta^2}{|\Delta^2-(\omega+i\eta)^2|}\right)\frac{f_T(V_s)}{\sqrt{1+\frac{\kappa_\omega^2}{\kappa_s^2}}}\frac{\cosh\left(\sqrt{\kappa_s^2+\kappa_\omega^2}(z-W)\right)}{\sinh\left(\sqrt{\kappa_s^2+\kappa_\omega^2}W\right)}.
\end{equation}
and $R_s=\frac{1}{\kappa_s\sigma_nA}$ is the normal state spin resistance of the barrier. This makes explicit that the perturbative expansion is controlled by the small ratio $R_s/R_b$.

For a superconducting wire formed into a closed loop of length $L$, the total circulating current along the $x$-direction is:
\begin{equation}
    I_{SH}=\frac{1}{L}\int_0^Ldx\int_0^Wdz\int_{-\infty}^\infty \left(-\frac{\sigma_n}{16eD}\right)\operatorname{Tr}_\sigma(\tau_3\mathcal{J}_x^K)d\omega=\frac{1}{L}\int_0^Ldx\int_0^Wdz\int_{-\infty}^\infty \frac{\theta\sigma_n}{8e}\operatorname{Tr} (\tau_3\partial_zg_{1t}^K)d\omega.
\end{equation}
Here we have exploited the spatial periodicity along the $x$-direction, such that all terms involving $\partial_x$ integrate to zero over a full period. Substituting Eq.~\eqref{eq:B16} yields:
\begin{equation}
    I_{SH}=-\theta\frac{\sigma_n}{e}\frac{L_{\mathrm{inj}}}{L}\int_0^\infty\frac{S_\omega(0)-S_\omega(W)}{N_0}d\omega
\end{equation}

\subsection{Spin to charge conversion}\label{B2}
Substituting Eqs.~\eqref{eq:gRA_expansion} and \eqref{eq:gK_expansion} into Eq.~\eqref{eq:Usadel_Keldysh}, retaining terms up to first order in $R_b^{-1}$, and taking the trace over spin space yields the equation for $g_{1s}^K$:
\begin{equation}\label{eq:B21}
    \nabla^2g_{1s}^K-\kappa_\omega^2 g_{1s}^K=0.
\end{equation}
The corresponding boundary conditions, derived from Eqs.~\eqref{eq:KLbcond_Keldysh1} and \eqref{eq:KLbcond_Keldysh2}, read:
\begin{equation}\label{eq:B22}
    \left.\partial_z g_{1s}^K\right|_{z=0}=\left.\theta g_0^R\partial_x g_{1t}^K\right|_{z=0}+\frac{1}{\sigma_nR_bA}(f_L(V_s)-f_L(0))(\tau_3+g_0^R\tau_3g_0^A) \Theta\left(\frac{L_{\mathrm{inj}}}{2}-|x|\right)
\end{equation}
\begin{equation}\label{eq:B23}
    \left.\partial_z g_{1s}^K\right|_{z=W}=\left.\theta g_0^R\partial_x g_{1t}^K\right|_{z=W}.
\end{equation}
The charge imbalance $Q^*$ is defined as:
\begin{equation}
    Q^*=\int_{0}^\infty Q^*_\omega d\omega=\frac{1}{4}\int_{0}^\infty \operatorname{Tr}\left(g^K_{1s}\right)d\omega.
\end{equation}
Taking the traces of Eqs.~\eqref{eq:B21}--\eqref{eq:B23} and substituting Eq.~\eqref{eq:B16}, we obtain the equations for $Q_\omega^*$:
\begin{equation}\label{eq:motionQ}
    \nabla^2Q_\omega^*-\kappa_\omega^2 Q_\omega^*=0
\end{equation}
\begin{equation}
    \left.\partial_zQ_\omega^*\right|_{z=0}=\theta q_\omega \frac{S_\omega(0)}{N_0} \partial_x\Theta\left(\frac{L_{\mathrm{inj}}}{2}-|x|\right)
\end{equation}
\begin{equation}
    \left.\partial_zQ_\omega^*\right|_{z=W}=\theta q_\omega \frac{S_\omega(W)}{N_0} \partial_x\Theta\left(\frac{L_{\mathrm{inj}}}{2}-|x|\right)
\end{equation}
where $q_\omega=\operatorname{Re}\left(\frac{\sqrt{(\omega+i\eta)^2-\Delta^2}}{\omega}\right)$ is the effective quasiparticle charge. Solving via separation of variables gives:
\begin{equation}
    Q_\omega^*=\theta\int_{-\infty}^\infty\frac{iq\Theta_q}{\sqrt{\kappa_\omega^2+q^2}}\left[\frac{q_\omega S_\omega(W)\cosh\left(\sqrt{\kappa_\omega^2+q^2}z\right)-q_\omega S_\omega(0)\cosh\left(\sqrt{\kappa_\omega^2+q^2}(z-W)\right)}{N_0\sinh\left(\sqrt{\kappa_\omega^2+q^2}W\right)}\right]\frac{e^{iqx}}{2\pi}dq.
\end{equation}
To simplify this expression, we integrate over $\omega$ and average over the width $W$ of the wire:
\begin{equation}
    \overline{ Q^*}=\frac{1}{W}\int_0^Wdz\int_{0}^\infty d\omega Q_\omega =\theta\int_0^\infty d\omega\frac{q_\omega(S_\omega(W)-S_\omega(0))}{N_0W}\int_{-\infty}^\infty\frac{iq\Theta_q}{\kappa_\omega^2+q^2}\frac{e^{iqx}}{2\pi}dq.
\end{equation}
Evaluating the Fourier integral in $q$ yields:
\begin{equation}
    \overline{Q^*}(x)=\theta\int_0^\infty d\omega\frac{q_\omega(S_\omega(0)-S_\omega(W))}{N_0\kappa_\omega W}\cdot
\begin{cases}
 -\sinh(\frac{\kappa_\omega L_{\mathrm{inj}}}{2})e^{\kappa_\omega x} & \text{if } x < -L_{\mathrm{inj}}/2 \\
\quad e^{-\kappa_\omega L_{\mathrm{inj}}/2}\sinh(\kappa_\omega x) & \text{if } -L_{\mathrm{inj}}/2\leq x\leq L_{\mathrm{inj}}/2 \\
\quad\sinh(\frac{\kappa_\omega L_{\mathrm{inj}}}{2})e^{-\kappa_\omega x} & \text{if } x > L_{\mathrm{inj}}/2 
\end{cases}.
\end{equation}
This expression demonstrates that, via the spin Hall angle $\theta$, the difference in spin accumulation between the top and bottom surfaces, $S_\omega(0) - S_\omega(W)$, acts as a source of charge imbalance.

The net charge imbalance drop between the two detector contacts shown in Fig.~\ref{fig:Setup}\textcolor{goodred}{(a)--(b)} is given by:
\begin{equation}\label{eq:chargeimbalancedrop}
    \delta Q^*=\overline{Q^*}(x_{\mathrm{det}})-\overline{Q^*}(-x_{\mathrm{det}})=2\theta\int_0^\infty d\omega \frac{q_\omega (S_\omega(0)-S_\omega(W))}{N_0\kappa_\omega W}\sinh\left(\frac{\kappa_\omega L_{\mathrm{inj}}}{2}\right)e^{-\kappa_\omega x_{\mathrm{det}}}.
\end{equation}

Experimentally, detector contacts are placed at distances $x_{\mathrm{det}} \gg \xi$, where $\xi$ is the superconducting coherence length. At these scales, the subgap contribution $|\omega| < \Delta$ is exponentially suppressed due to the energy dependence of $\kappa_\omega$:
\begin{equation}
\kappa_\omega^{-1}\sim \begin{cases} \xi & |\omega|<\Delta\\ \lambda_{Q^*} & |\omega|\geq\Delta \end{cases} 
\end{equation} 
where $\lambda_{Q^*}$ is the inelastic charge-imbalance relaxation length (typically on the order of several micrometers). Because $\lambda_{Q^*}\gg\xi$, the subgap contribution decays rapidly away from the injector and contributes negligibly to $\delta Q^*$ compared to the contribution with $|\omega| > \Delta$. Furthermore, the prefactor $q_\omega$ is also small subgap energies. Therefore, in general one can safely set the lower integration limit in Eq.~\eqref{eq:chargeimbalancedrop} to $\Delta$.

\subsection{\texorpdfstring{Charge imbalance in the limit $\eta\to0$}
{Charge imbalance in the limit η → 0}}\label{B3}
In the limit $\eta \to 0$, Eq.~\eqref{eq:chargeimbalancedrop} simplifies further. For $|\omega| > \Delta$, $\kappa_\omega\to0$ and the integrand simplifies via
\begin{equation}
    q_\omega\frac{S_\omega(0)-S_\omega(W)}{N_0}=\begin{cases}  0 & \text{if } |\omega|<\Delta\\ \frac{R_s}{R_b} N_\omega f_T(V_s)\tanh\left(\frac{\kappa_sW}{2}\right) & \text{if } |\omega|\geq\Delta \end{cases} 
\end{equation}
Substituting this relation into Eq.~\eqref{eq:chargeimbalancedrop} yields the final result:
\begin{equation}\label{eq:B35}
    \lim_{\eta\to0}\delta Q^*=r\int_\Delta^{\infty} N_\omega f_T(V_s)d\omega
\end{equation}
where
\begin{equation}
    r = \theta\frac{R_s}{R_b}\frac{L_{\mathrm{inj}}}{W}\tanh\left(\frac{\kappa_s W}{2}\right).
\end{equation}

\subsection{Detector signal}
The detector voltage $V_{\mathrm{det}}$ is determined by imposing the zero-current condition, yielding:
\begin{equation}
    \delta Q^*=\int_0^{\infty} N_\omega f_T(V_{\mathrm{det}})d\omega.
\end{equation}
Substituting Eq.~\eqref{eq:chargeimbalancedrop} gives an integral equation for $V_{\mathrm{det}}$:
\begin{equation}
    2\theta\int_0^\infty d\omega \frac{q_\omega (S_\omega(0)-S_\omega(W))}{N_0\kappa_\omega W}\sinh\left(\frac{\kappa_\omega L_{\mathrm{inj}}}{2}\right)e^{-\kappa_\omega x_{\mathrm{det}}}=\int_0^{\infty} N_\omega f_T(V_{\mathrm{det}})d\omega
\end{equation}
or substituting Eq.~\eqref{eq:B35} in the limit $\eta\to0$:
\begin{equation}
    r\int_\Delta^{\infty} N_\omega f_T(V_s)d\omega=\int_\Delta^{\infty} N_\omega f_T(V_{\mathrm{det}})d\omega.
\end{equation}
\end{document}